\documentstyle[prl,preprint,aps]{revtex}

\begin{document}
\tighten
\draft
\preprint{ }
\title{A microscopic T-violating optical potential:\\ Implications
for neutron-transmission experiments}
\author{J. Engel,$^{\rm a}$ C. R. Gould$^{\rm b}$ and V. Hnizdo$^{\rm c}$}
\address{$^{\rm a}$Department of Physics and Astronomy,
University of North Carolina, Chapel Hill, North Carolina 27599  }
\address{$^{\rm b}$Department of Physics, North Carolina State University,
 Raleigh, North Carolina 27695\\
and Triangle Universities Nuclear Laboratory, Durham, North Carolina 27708}
\address{$^{\rm c}$Department of Physics and Schonland Research Centre for
 Nuclear Sciences,
\\ University of the Witwatersrand, Johannesburg, 2050 South Africa}
\date{\today}
\maketitle
\begin{abstract}
We derive a T-violating P-conserving optical potential for neutron-nucleus
scattering, starting from a uniquely determined two-body $\rho$-exchange
interaction with the same symmetry. We then obtain
limits on the T-violating $\rho$-nucleon coupling $\overline{g}_{\rho}$ from
neutron-transmission experiments in $^{165}$Ho.  The limits may soon
compete with those from measurements of atomic electric-dipole moments.
\end{abstract} 

\narrowtext

Several experimental groups have recently scattered polarized neutrons with low
energies from heavy nuclei in search of parity or time-reversal
violation\cite{r:proceedings}.  The time-reversal experiments, requiring
oriented targets, are the more difficult to perform.  So far the only published
measurements of this kind\cite{r:Gould} have used neutron beams of several MeV
and aligned $^{165}$Ho targets to search for a dependence of the forward
elastic-scattering amplitude on the ``five-fold" T-violating P-conserving
(TVPC)
correlation term $(\hat{\bf s} \cdot \hat{\bf I} \times \hat{\bf p} ) (
\hat{\bf
I} \cdot\hat{ \bf p })$.  Here $\hat{\bf I}$ and $\hat{\bf s}$ are unit vectors
along the axes of the target alignment and neutron polarization, and $\hat{\bf
p}$ is the direction of the incident neutron beam.  The correlation in the
forward amplitude shows up as a change in the total cross section --- related
to
the amplitude through the optical theorem --- when the polarization of the
incident neutron spin is reversed.

Until now there has been no credible way to connect these experiments to
fundamental sources of T-violation.  Within the standard model TVPC effects
must be tiny, but in extended models this may not be the case.  Here we take
an important first step in using neutron-transmission experiments to test any
such models:  Through the construction of a microscopic T-violating optical
potential, we show how to relate the TVPC observable $A_5$\cite{r:Hnizdo}
(connected to the difference between total cross sections for spin-up
and spin-down neutrons on an aligned target) to TVPC meson-nucleon coupling
constants.  The problem of constraining fundamental models of T-violation then
reduces to physics at the meson/nucleon scale --- namely a description of the
effective TVPC vertex in terms of quarks and gluons.

In the case of parity violation, a number of mesons ($\pi$, $\rho$, $\omega$,
etc.)  contribute to the force between nucleons, which in turn determines the
optical potential.  Fortunately for us, the TVPC interaction is severely
constrained.  Simonius showed some time ago\cite{r:Simonius} that only the
$\rho^{\pm}$ and $A_1$ (or heavier) mesons can contribute at tree level;
one-pion exchange, for example, is not allowed.  Moreover, the form of the
$\rho$-exchange potential is unique.  The $A_1$-exchange force is less
constrained, but the $A_1$ is significantly heavier than the $\rho$, and so its
effects are damped by short-range nucleon-nucleon repulsion.  We therefore
restrict our attention to $\rho$-exchange; the techniques we use can just as
easily be applied to the $A_1$.

The unique TVPC $\rho$-exchange interaction is\cite{r:Simonius,r:Haxton}
\begin{eqnarray}
\label{e:exchange}
 V^{\rho}_{1,2} & =& {\cal V}^{\rho}_{1,2} ~ [\tau_1 \times \tau_2 ]_3  \\
{\cal V}^{\rho}_{1,2} & = &{m_{\rho}^3   g_{\rho}^2
\overline{g}_{\rho} ~ \mu_v \over 4 \pi M^2} ~ {e^{-m_{\rho}r_{12}} \over
m_{\rho}^3 r_{12}^3 } (1 + m_{\rho} r_{12} ) (\mbox{\boldmath $\sigma$}_1 -
\mbox{\boldmath $\sigma$}_2 ) \cdot {\bf l} ~ , \nonumber
\end{eqnarray}
where ${\bf r}_{12} = {\bf r}_1 - {\bf r}_2$, ${\bf l} = {\bf r}_{12} \times
\frac{1}{2} ({\bf p}_1 - {\bf p}_2)$, $\mu_v = 3.70$ n.m.\ is the isovector
nucleon magnetic moment, $M$ is the nucleon mass, $g_{\rho}=2.79$ is the
normal strong $\rho NN$ coupling, and $\overline{g}_{\rho}$ is a
dimensionless ratio of the TVPC coupling to $g_{\rho}$.  This notation is
taken from Ref.\ \cite{r:Haxton}, where limits on $\overline{g}_{\rho}$ were
deduced from limits on the electric dipole moments of the neutron and
${}^{199}$Hg.

The potential in Eq.\ (\ref{e:exchange}) has a number of
peculiarities.  The isospin piece $[\tau_1 \times \tau_2 ]_3$ = $2i(\tau_1^+
\tau_2^- - \tau_1^- \tau_2^+ )$, which is solely responsible for the
T-violation through the factor $i$, implies that neutrons interact only with
protons and that the direct part of any two-nucleon matrix element vanishes.
In addition, the space-spin part of the interaction, while hermitian, is
antisymmetric in the nucleon coordinates.  Combined with the spin-dependence
of the force, these features have important consequences for the strength of
the neutron-nucleus TVPC optical potential $\bar{U}({\bf r})$, which we now
proceed to evaluate.

We begin by considering a closely related quantity $\bar{M}({\bf r})$,
the ground-state expectation value of the interaction
$V^{\rho}_{1,2}$:
\begin{equation}
\label{e:definition}
\langle \psi_a | \bar{M} |  \psi_b \rangle
\equiv \langle \psi_a \Psi | \sum_p V^{\rho}_{1,p} | \psi_b
 \Psi \rangle_A ~,
\end{equation}
where the index 1 refers to the incident neutron, $| \psi_a\rangle$ and
$|\psi_b \rangle$ are arbitrary neutrons states, the index $p = 2,3,...Z+1$
labels protons in the nucleus, $| \Psi \rangle$ is the nuclear ground state,
and the
subscript $A$ means that the matrix element is completely antisymmetrized.
The optical potential itself, $\bar{U}({\bf r})$ --- in the so-called
``adiabatic approximation"\cite{r:Feshbach} --- is a function of the
nuclear-spin operator ${\bf I}$, the neutron-spin operator ${\bf s} =
\frac{1}{2} \mbox{\boldmath $\sigma$}$, and the neutron position ${\bf r}$,
defined so that
the ground-state nuclear matrix element of $\bar{U}$ is $\bar{M}$.  This
definition, which amounts to treating the nucleus as an elementary particle
with spin $I$, is necessary for making contact with the usual phenomenology,
where terms that depend on the nuclear spin (e.g.\ $\mbox{\boldmath $\sigma$}
\cdot {\bf I}$) appear in the optical potential.

A number of authors have derived
similar one-body potentials --- usually felt by $bound$ nucleons\cite{r:Vogel}
--- in the
approximation just described.  Unfortunately, the unusual features of
$V^{\rho}$ make the same kind of calculations more complicated here.  If the
nuclear wave function is approximated as  usual by a Slater determinant,
then
\begin{equation}
\label{e:HF}
\langle {\bf r}_1 | \bar{M} | {\bf r}_2  \rangle =
 2 i
\sum_p \langle {\bf r}_1 | \langle
\phi_p | ~{\cal V}^{\rho}_{1,2}~ | \phi_p  \rangle | {\bf r}_2
\rangle ,
\end{equation}
where the $| \phi_p \rangle$ are occupied proton orbits.  Eq.\ (\ref{e:HF})
sums only {\it exchange} matrix elements in coordinate and spin space; the
optical potential is therefore entirely nonlocal.  Clearly the ``folding'' of
potential and nuclear density that is often used to construct ordinary optical
potentials is not feasible.  Furthermore, the various
methods\cite{r:nonlocal} for constructing equivalent local potentials
all make simplifying assumptions that do not apply here.  Luckily, the
${\rho}$ is heavy enough that a zero-range approximation will be accurate
provided strong $NN$ repulsion is temporarily ignored.  We
therefore rewrite ${\cal V}^{\rho}$ as
\begin{equation}
\label{e:zerorange}
\FL
{\cal V}^{\rho}_{1,2} = - {g_{\rho}^2
\overline{g}_{\rho} ~ \mu_v \over 4 \pi M^2} ~ {\bf \nabla}_{12}
\left( {
e^{-m_{\rho} r_{12}} \over r_{12}} \right) \times {\bf p}_{12}
\cdot
(\mbox{\boldmath $\sigma$}_1 - \mbox{\boldmath $\sigma$}_2 ) ,
\end{equation}
where ${\bf \nabla}_{12}
= \frac{1}{2} ({\bf \nabla}_1 - {\bf \nabla}_2 )$,
and ${\bf p}_{12} = -i
{\bf \nabla}_{12}$.  We then take the limit $m_{\rho}
\rightarrow \infty$, which amounts to replacing the quantity in parentheses by
a delta function of ${\bf r}_{12}$.
The sum in Eq.\ (\ref{e:HF}) will now result in
a local potential.  To obtain it, we first evaluate $\langle \psi_a |
\bar{M} | \psi_b \rangle$ in the delta-function limit by
going to relative and CM coordinates and integrating by
parts, yielding
\widetext
\begin{equation}
\label{e:intermediate}
\FL
\langle \psi_a |\bar{M} | \psi_b \rangle \approx {g_{\rho}^2
\overline{g}_{\rho}
\mu_v \over m_{\rho}^2 M^2} \sum_p \int d^3r ~ \psi_a^{\dag}({\bf r})
\bigg(
\big[ ~\mbox{\boldmath $\sigma$} ~ \cdot , {\bf \nabla} \phi_p({\bf r})
           \times {\bf \nabla} \phi_p^{\dag}({\bf r}) ~ \big]
                      - {\bf \nabla} \times  \big[ ~\mbox{\boldmath $\sigma$} ,
\phi_p({\bf r}) \phi_p^{\dag}({\bf r}) ~\big] \cdot  {\bf \nabla} \bigg)
\psi_b({\bf r}) ,
\end{equation}
\narrowtext
\noindent
where the derivatives act only on the functions immediately next to them, and
the square brackets
indicate commutators, combined in the first term with a dot product.
$\bar{M}({\bf r})$ is now just the operator in parentheses
above.  Noting that the $\phi_p
({\bf r})$ are two-component spinors ($\phi \phi^{\dag}$ is an
outer product), we use trace
identities to obtain
\begin{equation}
\label{e:final}
\FL
\bar{M}({\bf r}) \approx {2 g_{\rho}^2
\overline{g}_{\rho} \mu_v \over m_{\rho}^2 M^2} \! \sum_{p,k} {\rm Re} \big(
 \,  i  \nabla_k \phi_p^{\dag}({\bf r}) \,
\mbox{\boldmath $\sigma$} \cdot
{\bf \nabla} \phi_p({\bf r}) \, \big)\,  \sigma_k,
\end{equation}
where $k$ labels Cartesian components.  We have dropped two other terms,
coming from the second term in Eq.\ (\ref{e:intermediate}), which change
sign
when the nuclear spin is reversed.  Since the
five-fold correlation itself is invariant under reversal of the nuclear spin
direction, the neglected terms can contribute only in conjunction
with a potential like $\mbox{\boldmath $\sigma$} \cdot {\bf I}$,
which rectifies their spin-reversal behavior.  Such terms
do exist
in the normal optical potential\cite{r:Thompson}, but are substantially
weaker than the central potential.  If we imagine treating them together with
our TVPC potential as perturbative corrections in DWBA, it is clear that the
term we have kept in Eq.\ (\ref{e:final}), which already has the right
symmetry,
will dominate those we have omitted.

Some of the potential's unusual
properties follow directly from Eq.\ (\ref{e:final}).  If, for example,
the $\phi_p({\bf r})$ come from a spherical potential, then
$\bar{M}$ and $\bar{U}$
vanish in a spin-saturated or closed-j-shell nucleus.  Thus only a few
valence protons in the last orbital contribute to
$\bar{U}$.  A spherical mean-field treatment in fact makes sense only in
closed-shell or closed-shell-plus-one nuclei, so that in this simple picture
at most one nucleon, characterized, e.g., by a valence orbital with quantum
numbers
$n,l,j$, is relevant.  Using standard angular-momentum algebra,
properties of spherical harmonics, etc., one can write down
an explicit expression for the $\bar{M}$ generated by the valence
orbital, and then for the corresponding optical potential $\bar{U}$.  For
$j= l \pm 1/2$ we have
\begin{equation}
\label{e:spherical}
 \bar{U}({\bf r}) ={5 ~\pi^{-1} \sqrt{30}~\hat{j}~  (-1)^{j-1/2} ~
g_{\rho}^2 ~
\overline{g}_{\rho} ~ \mu_v \over \sqrt{ j (j+1)(2j-1)(2j+3)}  m_{\rho}^2
 M^2}  Z_{nlj,nlj}(r)~T_5,
\end{equation}
where $\hat{a} \equiv \sqrt{2a+1}$ and $Z_{nlj,nlj}(r)$ is the diagonal
component
of a matrix we will use again below ($j'=l' \pm 1/2$):
\widetext
\begin{eqnarray}
\label{e:Z}
&&Z_{nlj,n'l'j'}(r) = \sqrt{2l' +1 \pm 2}
    \left\{ \begin{array}{ccc}
              2    & l & l' \pm 1 \\
              1/2  & j' & j
             \end{array} \right\}
    \left[   \sqrt{(l+1)(2l+1)}  \left\{ \begin{array}{ccc}
                                           2   & l & l' \pm 1 \\
                                           l+1 & 2 & 1
                                          \end{array} \right\}
\makebox[1.2in]{}
 \right.\\
        && \left.   \times          \left(  \begin{array}{ccc}
                                           l+1 & l' \pm 1 & 2 \\
                                            0  &    0    & 0
                                           \end{array} \right)
                                        R^+_{nlj}(r)
                 ~   - ~ \sqrt{l(2l-1)}   \left\{ \begin{array}{ccc}
                                           2   & l & l' \pm 1 \\
                                           l-1 & 2 & 1
                                          \end{array} \right\}
                                  \left(  \begin{array}{ccc}
                                           l-1 & l' \pm 1 & 2 \\
                                            0  &    0    & 0
                                          \end{array}  \right)
                                       R^-_{nlj}(r) \right]
                   R^{\pm}_{n'l'j'}(r) ~.\nonumber
\end{eqnarray}
\narrowtext
\noindent
The $R^{\pm}_{nlj}$ are related to the radial wave functions $R_{nlj}$ by
\begin{equation}
\label{e:radial}
R_{nlj}^{\pm}(r) \equiv  \left(\frac{\partial}{\partial r} \mp \frac{l + 1/2
\mp 1/2}{r} \right) R_{nlj}(r)~.
\end{equation}

The factor $T_5$ in Eq.\ (\ref{e:spherical}) is the ``five-fold" operator
\begin{eqnarray}
T_5&=&\case{1}{2} r^{-2}\big({\bf s}\cdot({\bf I}\times {\bf r})
({\bf I}\cdot {\bf r})+({\bf I}\cdot
 {\bf r})({\bf I}\times {\bf r})\cdot{\bf s}\big)\nonumber \\
&=&-i\sqrt{\pi}~\big[[{\bf I} \times {\bf I}]^2 \times [Y_2(\hat{\bf r}) \times
\mbox{\boldmath $\sigma$}]^2\big]^0~,
\label{e:T5}
\end{eqnarray}
where the square brackets now mean that the angular momenta are
coupled. (We have omitted from Eq. (\ref{e:spherical})
other terms that have no effect when
the target is only rank-2 aligned.)
The presence of $T_5$ in Eq.\ (\ref{e:spherical}) is interesting;
it was proposed without microscopic justification in Ref.\ \cite{r:Hnizdo}
solely because of its TVPC tensorial structure.  The {\it radial\/ } form of
Eq.\ (\ref{e:spherical}), however, is quite different from the Wood-Saxon
used in Ref.\ \cite{r:Hnizdo}, a point to which we return shortly.

First, however, having calculated  $\bar{U}$ for spherical single-particle
states, we address ${}^{165}$Ho, the only nucleus in
which the five-fold correlation has actually been measured.  To incorporate
nuclear deformation, we use the Nilsson model with $\epsilon = 0.3$.  We
repeat the evaluation of the mean field potential in Eq.\
(\ref{e:final}) --- this time in the intrinsic frame --- ignoring the weak
dependence of the TVPC hamiltonian on the Euler angles.  Next we expand the
intrinsic state in terms of spherical
states (neglecting admixtures with $\Delta N = 2$)
and integrate over Euler angles in the usual way\cite{r:Deshalit}.
With some rearrangement of terms, we arrive at the expression
\widetext
\begin{eqnarray}
\label{e:Nilexp}
\bar{U}({\bf r})& = & {10 ~ \pi^{-1} \sqrt{30 } ~ \hat{I}~  (-1)^{I+K} ~
g_{\rho}^2 ~ \overline{g}_{\rho} ~ \mu_v \over  \sqrt{ I (I+1)(2I-1)(2I+3)}
  ~ m_{\rho}^2  ~ M^2}
                        \left( \begin{array}{ccc}
                               I  &  I & 2 \\
                               -K  & K & 0
                       \end{array} \right)  \\
&\times& \bigg[~\sum_{nlj,n'l'j',\Omega >0}^{}
\hat{j} ~ \hat{j'} \, (-1)^{1/2+\Omega} ~ a^{\Omega}_{nlj} \,
a^{\Omega}_{n'l'j'}   \left( \begin{array}{ccc}
			      j'       &  j     &  2 \\
			      - \Omega & \Omega &  0
                        \end{array} \right)
Z_{nlj,n'l'j'}(r)~ \bigg] ~   T_5   , \nonumber
\end{eqnarray}
\narrowtext
\noindent
where $I=7/2$ is the nuclear spin, $K=7/2$ is the z-projection in the intrinsic
frame, and the $a$'s are spherical expansion coefficients for the deformed
single-particle orbital labeled by $\Omega$.  Even though all such orbits in
the valence shell now contribute to $\bar{U}$, in the end the strength of
the potential is still comparable to that arising from a single spherical
orbital.  Figure~\ref{f:1} shows the potential's radial shape.  The curve
looks very different from a typical volume or
surface optical potential, reflecting the nonlocality discussed above. It is
small both at the origin and the surface, peaking somewhere in between.

Including the TVPC potential alongside the strong optical potential
$U$\cite{r:Hnizdo} in, e.g., the coupled-channels code CHUCK\cite{r:Kunz}, we
can calculate the spin-correlation coefficient $A_5$ for any value of
$\overline{g}_{\rho}$, or vice versa.  Figure~\ref{f:2} shows $A_5$
as a function of neutron-energy for $\overline{g}_{\rho} = 1$.
The published measurement of $A_5$ at 2 MeV\cite{r:Gould} results in an upper
limit on $\overline{g}_{\rho}$ of about 0.5, a value that must be increased by
a factor of about 3 to account for short-range
repulsion\cite{r:Haxton}, which we have so far neglected.  The additional
analysis in Ref.\ \cite{r:Haxton} then implies a limit on $\alpha_T$, the
ratio of typical T-violating to strong two-body matrix elements, of about $1.5
\times 10^{-2}$.  A recently completed experiment has improved the bound on
$A_5$ by a factor of about 15, however\cite{r:Gould1}, and a further order of
magnitude is anticipated, potentially resulting in bounds on $\alpha_T$ of
order $10^{-4}$; this would make neutron-transmission experiments competitive
with measurements of atomic dipole moments.  Ref.\ \cite{r:Haxton} also
derives another limit, roughly an order of magnitude smaller still, from
measurements of the neutron dipole moment, but that value depends on
the parity-violating pion-nucleon coupling, the size of which has been
estimated\cite{r:DDH} but is not known reliably.

In Ref.\ \cite{r:Hnizdo}, a Wood-Saxon shape was used for the radial potential
multiplying $T_5$
under the assumption that the potential is proportional to the nuclear
density.  A limit was then obtained on $\alpha_T$ by dividing the strength of
the phenomenological TVPC potential (determined by calculating $A_5$ in the
same way as is done here) by that of the central optical potential.  Our
results show that this procedure yields too small a limit by about two orders
of magnitude.  Our TVPC potential is generated in lowest order by at most a
few nucleons in the valence shell --- hence the different shape and
considerably larger upper limit on $\alpha_T$.  A more appropriate way to
extract a rough limit from a phenomenological potential would be to divide its
strength not by that of the central potential, but instead by the strength of
a symmetry-conserving spin-spin term such as $\big[ [{\bf I} \times {\bf I}]^2
\times [{\bf l} \times \mbox{\boldmath $\sigma$}]^2 \big]^0$, which resembles
our TVPC interaction.  In general, spin-spin interactions are $\approx 100$
times weaker than central interactions because they too are generated only by
valence nucleons.

How reliable are the present results?  The adiabatic approximation of Eq.\
(\ref{e:definition}) ignores higher-order (in $V_{\rm strong}$) processes in
which intermediate nuclear or neutron states are virtually excited and
deexcited.  Although such terms would alter our potential, we do not expect
extremely large changes.  Any coherent (higher-order) one-body potentials must
be independent of the nuclear spin $I$, and therefore can affect $I$-dependent
correlations only when acting together in perturbation theory with some other
$ I$-dependent force.  Higher-order contributions to $\bar{U}$ that are
themselves $I$-dependent should be suppressed by amounts typical of
Bruckner-Bethe-Goldstone perturbation theory.  Additionally, working in a
deformed basis takes into account at least some of the important nuclear
correlations.  Our potential is therefore probably correct to within factors
of order unity.

Several times we have noted that T-odd potentials generated by the core will
not contribute by themselves to $I$-dependent observables in lowest order.
This statement, which is true whether the potentials are P-even or P-odd, does
not however appear to be relevant for epithermal neutrons because in a compound
nucleus $I$-dependent terms can no longer be treated as perturbative
corrections.  On the other hand, because of the complicated structure of
compound-nucleus resonances, there is no simple connection of
the kind established here between experimental observables and T-odd forces.
It therefore remains to be seen whether experiments with
epithermal neutrons can constrain TVPC couplings as reliably as experiments at
higher energies.

We wish to acknowledge useful conversations with A.  H\"{o}ring and especially
with W.C.  Haxton.  J.E.\ thanks the Institute for Nuclear Theory,
and V.H.\ the Triangle Universities Nuclear Laboratory for their hospitality
while parts of this work were carried out.  We were supported in part by the
U.S.  Department of Energy under grants DE-FG05-94ER40827 and
DE-FG05-88ER40441.

\begin{figure}
\caption{The radial part of the TVPC optical potential $\bar{U}$ (with
$\overline{g}_{\rho} = 1$) multiplying $T_5$ in Eq.\ (\protect\ref{e:Nilexp}).}
\label{f:1}
\end{figure}

\begin{figure}
\caption{The spin-correlation coefficient $A_5$ \protect\cite{r:Hnizdo} as a
function of neutron energy for
$\overline{g}_{\rho} = 1$.}
\label{f:2}
\end{figure}

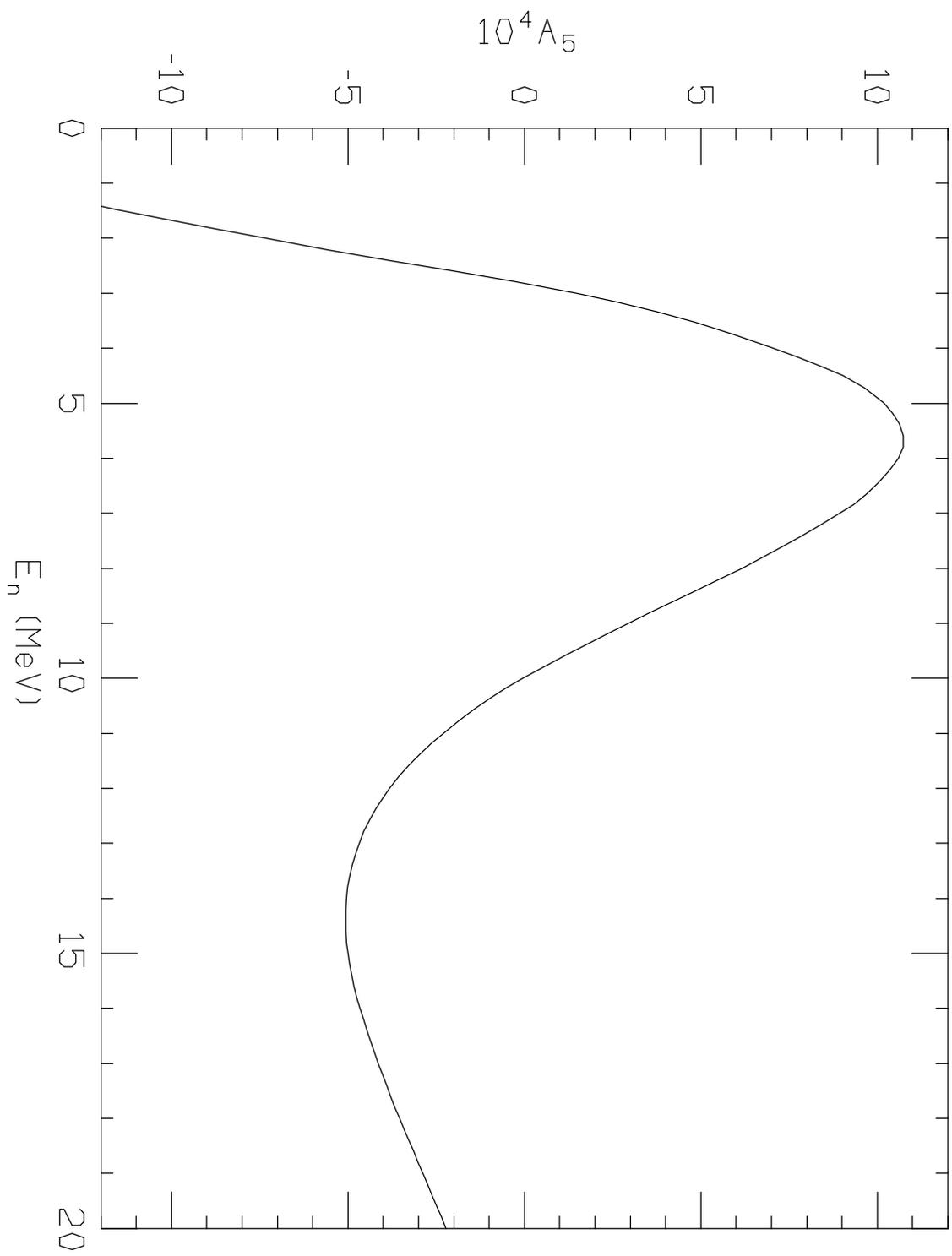
\end{document}